\documentclass[12pt, a4paper]{article}

\makeatletter
\renewcommand\footnotesize{%
   \@setfontsize\footnotesize\@xipt\@xipt
   \abovedisplayskip 10\p@ \@plus2\p@ \@minus5\p@
   \abovedisplayshortskip \z@ \@plus3\p@
   \belowdisplayshortskip 6\p@ \@plus3\p@ \@minus3\p@
   \def\@listi{\leftmargin\leftmargini
               \topsep 6\p@ \@plus2\p@ \@minus2\p@
               \parsep 3\p@ \@plus2\p@ \@minus\p@
               \itemsep \parsep}%
   \belowdisplayskip \abovedisplayskip
}
\makeatother

\usepackage[american]{babel}

\usepackage[babel]{csquotes}

\usepackage[authordate, natbib, isbn=false, backend=biber, noibid]{biblatex-chicago}  
\usepackage{authblk}

\usepackage[bottom]{footmisc}

\usepackage{amsmath}
\usepackage{amsfonts}

\usepackage{caption}
\usepackage{geometry} 
\usepackage{setspace}
\usepackage{etoolbox}
\newcommand{\zerodisplayskips}{%
  \setlength{\abovedisplayskip}{0pt}%
  \setlength{\belowdisplayskip}{12pt}%
  \setlength{\abovedisplayshortskip}{0pt}%
  \setlength{\belowdisplayshortskip}{12pt}}
\appto{\normalsize}{\zerodisplayskips}
\appto{\small}{\zerodisplayskips}
\appto{\footnotesize}{\zerodisplayskips}

\usepackage{graphicx}

\usepackage{booktabs} 

\usepackage{array}

\usepackage{paralist} 

\usepackage{verbatim}

\usepackage{subfig} 

\usepackage{fancyhdr} 
\usepackage[nottoc, notlof, notlot]{tocbibind} 
\usepackage[titles, subfigure]{tocloft} 

\begin{document}

\thispagestyle{empty}

\bigskip

\begin{center}
{\Large Which Type of Statistical Uncertainty Helps Evidence-Based Policymaking? An Insight from a Survey Experiment in Ireland}
\end{center}

\bigskip

\begin{center}
{\large Akisato Suzuki}

\begin{singlespace}
School of Politics and International Relations\\
University College Dublin\\
Belfield, Dublin 4, Ireland\\
akisato.suzuki@gmail.com\\
ORCID: 0000-0003-3691-0236
\end{singlespace}
\end{center}

\bigskip

\begin{center}
{\large Working paper\\ (February 11, 2024)}
\end{center}
\begin{singlespace}

\begin{center}
\textbf{Abstract}
\end{center}

\noindent
Which type of statistical uncertainty -- statistical (in)significance with a \textit{p}-value, or a Bayesian probability -- enables people to see the continuous nature of uncertainty more clearly in a policymaking context? An original survey experiment used a hypothetical scenario, where participants from Ireland were asked whether to introduce a new bus line to reduce traffic jams, given a research report estimating its effectiveness. The treatments were uncertainty information: statistical significance with a \textit{p}-value of 2\%, statistical insignificance with a \textit{p}-value of 25\%, the 95\% probability that the estimate is correct, and the 68\% probability that the estimate is correct. In the case of lower uncertainty, both significance and Bayesian frameworks resulted in a large proportion of participants adopting the policy (0.82 and 0.91 respectively). In the case of higher uncertainty, the significance framework led a much smaller proportion of participants to adopt the policy (0.39 against 0.83). The findings suggest participants saw the continuous nature of uncertainty more clearly in the Bayesian framework than in the significance framework.
\end{singlespace}


\clearpage

\section{Introduction}
To evaluate the uncertainty of statistical estimates, the dichotomy of statistical significance vs. insignificance based on a \textit{p}-value has been a dominant approach, while Bayesian posterior probability has been proposed as a major alternative more recently \autocite{Kruschke2018}. The question this article addresses is: Which type of statistical uncertainty enables people to see the continuous nature of uncertainty more clearly, specifically in a policymaking context?

Uncertainty is a continuous scale, as most typically seen in a weather forecast. Rational policy decision making must consider how likely a policy is to produce a desired outcome. If research is to help evidence-based policymaking, it must communicate the degree of uncertainty such that non-specialists can understand it. Evidence-based/informed policymaking is an important agenda in international organizations and national governments to improve the quality of policymaking \autocite{OECD2019, ScienceAdviceforPolicybyEuropeanAcademies2019}.\footnote{In this article, I use ``evidence-based'' throughout to indicate policymaking either based on, or informed by, scientific evidence, as it is more common than ``evidence-informed'' according to Google (about 293,000 results for ``evidence-based policymaking'' while about 24,900 results for ``evidence-informed policymaking'' on January 15th, 2020). In the literature, there is a difference in the nuance; ``evidence-based'' implies a more dominant influence of scientific evidence on policymaking \autocite[22--23]{ScienceAdviceforPolicybyEuropeanAcademies2019}.} 

To address the aforementioned question, I ran a survey experiment on a sample from the population of Ireland and obtained 517 responses. The participants were asked to assume they were a policymaker. They were shown the interval estimate of the effect of a new bus line on a traffic jam reduction in a hypothetical scenario. They were asked to decide about whether to introduce the new bus line or not.

They were randomly assigned into one of the four groups where different treatments were given on uncertainty information. The first treatment was lower uncertainty under the significance framework: ``the estimate is statistically significant (the \textit{p}-value is 2\%),'' while the second was a higher uncertainty level: ``the estimate is not statistically significant (the \textit{p}-value is 25\%).''\footnote{At the preregistration stage, I used the term ``Frequentist'' framework instead of the ``significance'' framework. Since frequentist inference is more than significance testing \autocite{Lew2012}, I refer to it as the significance framework in this article.}  The third treatment was lower uncertainty under the Bayesian framework: ``the estimate is correct with a probability of 95\%,'' while the fourth was higher uncertainty: ``the estimate is correct with a probability of 68\%.'' The cost and effect size of the new bus line were fixed across all treatment groups.

The experiment measured  the decision the participants would make, given the different types and degrees of statistical uncertainty. The experiment was not about whether participants would choose a ``correct'' answer under specific information of uncertainty. Indeed, there was no correct answer in the experiment, as what is the best decision given a cost, an effect size, and the uncertainty of an estimate depends on each person's utility function. Rather, it was about whether participants' decision would differ between the significance and Bayesian frameworks and, if so, which of the two frameworks enabled participants to \textit{see the continuous nature of uncertainty}. On average, participants should be less likely to introduce the new bus line in the higher-uncertainty scenarios. It should be only \textit{less likely} rather than not at all, because there should be some who think the policy is worth trying even in the higher-uncertainty scenarios. Similarly, even in the lower-uncertainty scenario, it is possible that some decide not to adopt the policy if they are too risk-averse or find the policy not cost-effective enough.

Yet, participants under the significance framework might tend to ignore the continuous nature of a uncertainty measure such as a \textit{p}-value, and instead might deterministically think ``statistically significant'' means evidence for the effect while ``not statistically significant'' means evidence for the null effect. Meanwhile, under the Bayesian framework, participants might be more able to see the continuous nature of uncertainty, as probability is an intuitive continuous measure of uncertainty used in everyday life (e.g., in weather forecasts). The implication is that participants under the significance framework should exhibit a greater difference in the likelihood of adopting the policy between the lower-uncertainty and higher-uncertainty treatments than those under the Bayesian framework. This article empirically examines how large the difference could be.

Experiments by \textcite{McShane2016, McShane2017} found that researchers were often misled by the dichotomy of statistical significance vs. insignificance to consider statistical insignificance as evidence for the null effect. The question is then whether non-experts such as the general public and policymakers would do that as well. \textcite{McShane2016} offered suggestive evidence: when asked to describe (rather than make inference over) the difference in life expectancy between two different treatment groups, statistically untrained undergraduate students, on average, thought that there was a difference even when the \textit{p}-value shown was greater than 5\%. Given that their experiment showed a \textit{p}-value without explaining what it means, it remains the question whether those statistically untrained students actually ignored the \textit{p}-values shown in the vignette, rather than interpreting them as a continuous measure of uncertainty.  \textcite{McShane2016} also examined whether showing not only a \textit{p}-value but also the corresponding Bayesian posterior probability would help participants avoid the dichotomous thinking, which turned out to be the case. The differences between this article and \textcite{McShane2016} are (1) that my experiment provided an explanation of the \textit{p}-value, and (2) that it presented a Bayesian probability independently rather than together with the corresponding \textit{p}-value.

The findings of this article are summarized as follows. In the case of the lower-uncertainty treatments, both significance and Bayesian frameworks resulted in a large proportion of participants adopting the policy (0.82 and 0.91 respectively). In the case of the higher-uncertainty treatments, the significance framework led a much smaller proportion of participants to adopt the policy (0.39 against 0.83) than the Bayesian framework. These findings suggest participants saw the continuous nature of uncertainty more clearly under the Bayesian framework than under the significance framework. This conclusion was reached, given (1) the experimental setup that the p-value and the Bayesian probability per degree of uncertainty (lower vs. higher) were defined so that they mathematically corresponded to each other, and (2) the existing behavioral literature that has found that a continuous, numerical probability scale enables people to see and use the measure of uncertainty more accurately than a discrete, verbal scale  \autocites[e.g.,][]{Mandel2021, McShane2016, McShane2017, Mislavsky2021, Friedman2018}.

The remainder of the article is structured as follows. First, it explains the design of the survey experiment. This is followed by the description of statistical modeling and the discussion of the results. Finally, concluding remarks are stated.

\section{Design of the Survey Experiment}
\subsection{Setup}
I preregistered the design of the survey experiment at the OSF Registries \autocite{CenterforOpenScience2020} -- the preregistered information is available at \textcite{Suzuki2020e}. The full vignette of the survey is available in the supplemental document. The survey was conducted online via the survey company Qualtrics. The data were collected between August 20th and September 1st in 2020 from the population of Ireland.\footnote{More precisely, the responses were from those who confirmed they were Irish citizens. Qualtrics used its own panel from Ireland. Because the experiment was online and anonymous, however, it was impossible to objectively verify the citizenship of all participants.} The survey was unlikely to obtain a representative sample of the population of Ireland, because it was conducted online but 9\% of households was estimated to have no Internet access in 2019 according to the Central Statistics Office (CSO) of Ireland \autocite{CentralStatisticsOffice2020}.

The original sampling plan, as in the preregistration, was to recruit 500 participants from Qualtrics’ panel of a sample from the population of Ireland, with the aims of the 50\%-50\% male-female ratio and of each of the three age groups (18--30, 31--45, and above 45) being 33\%. However, after the preregistration, I adjusted these categories based on my actual demographic questions: an age-group question (18--24, 25--44, 45--64, 65 or above) based on age-group categories used in the population estimate by the CSO \autocite{CentralStatisticsOffice2021}, and a three-category gender question (male, female, and other/non-binary).

During the preregistration stage, I conducted power analysis, i.e., computed the rate of false negatives via Monte Carlo analysis, to check whether the sample size was large enough. The parameter values were set at those I used to specify informative priors for my Bayesian logistic regression model (see the section ``Statistical Modeling'' for the detail of these informative priors). The false negative rate was 0.999 in 1,000 iterations given the one-tailed 95\% credible level. This meant that the sample size of 500 was large enough to detect the treatment effect according to the conventional threshold. The detail of the power analysis is available in the supplemental document.

The cross-tab of gender and age groups in the obtained sample is presented in Table \ref{crosstab}, together with the 2020 population estimate by the CSO \autocite{CentralStatisticsOffice2021}. The sample contains more female participants in the 18--24 age group and more male participants in the 65+ age group than in the population estimate. In the analysis, I do estimation with and without regression adjustment for these demographic variables.

\begin{table}[t]
\centering
\begin{tabular}{l c c c c c c c c}
\hline
\hline
 & \multicolumn{4}{c}{sample} & \multicolumn{4}{c}{CSO population estimate} \\
 & 18--24 & 25--44 & 45--64 & 65+ & 15--24 & 25--44 & 45--64 & 65+ \\ 
\hline
Male & 29\% & 44\% & 51\% & 68\% & 51\% & 49\% & 50\% & 47\% \\
Female  & 67\% & 56\% & 49\% & 31\% & 49\% & 51\% & 50\% & 53\% \\
Other/Non-binary & 3\%  & 0\%  & 0\%  & 0\% & & & & \\
missing & 1\%  & 0\%  & 1\%  & 1\% & & & & \\
\hline
\hline
\end{tabular}
\caption{Cross-tab of gender and age groups. The columns do not necessarily sum to one because of rounding. Note that while the age group of young adults in the CSO include from 15 to 24 years old, my survey targeted those who were at least 18 years old.}
\label{crosstab}
\end{table}

Participants' attentiveness to the survey was checked based on whether participants met the following criteria: (1) answer the question to measure the outcome (the ``outcome question'' hereafter); (2) answer two manipulation checks correctly (for details on these checks, see the section ``Manipulation Checks'' in the supplemental document); (3) complete the survey not too quickly (at least 30 seconds; see the section ``Manipulation Checks'' in the supplemental document). I removed the participants who did not meet at least one of these criteria, to make my entire sample consist of attentive participants only. The section ``Manipulation Checks'' in the supplemental document discusses the possibility of post-treatment selection bias \autocite{Montgomery2018b}. Although it is another interesting topic to examine how inattentive people make a decision given different types of statistical uncertainty, this article focuses on how attentive people do so. A science communication study suggests that ``the presentation format hardly has an impact on people in situations where they have time, motivation, and cognitive capacity to process information systematically'' \autocite[284]{Visschers2009}. My experiment can investigate whether this is indeed the case.

The survey obtained total 517 responses that met the aforementioned criteria. In the supplemental document, I discuss (1) the possibility of the same participant having completed the experiment more than once, (2) to what extent this possibility could have affected my analysis, and (3) how I checked this possibility. My conclusion is, in summary, that it seems unlikely that the data contain those who were able to automatically provide random, poor-quality answers, and bias the estimation of an average treatment effect, even if there had been participants who took the experiment more than once.

A small compensation was given to those who completed the survey appropriately. The survey was anonymous, and no question was mandatory. The study was approved as an ethics review exemption by the University College Dublin Office of Research Ethics in advance; the exemption was appropriate because of the low-risk nature of the survey. I obtained an informed consent from all participants and also the confirmation that they are at least 18 years old and an Irish citizen. After agreeing on the informed consent form and giving this confirmation, participants proceeded to the first page of the questionnaire, where they were asked to answer demographic questions: gender (female, male, and other/non-binary), age groups (18--24, 25--44, 45--64, and 65 or above), whether they hold a university degree of Bachelor’s or higher (yes or no), and whether they have studied or used statistics previously (yes or no).

After this, the participants were randomly assigned to one of the four treatment conditions. The randomization was set up such that all treatment conditions should have received the same number of participants. However, the exactly same number per treatment could not be achieved because there was the variation in the number of participants who satisfied the aforementioned necessary criteria to be included in the data. The number of responses per treatment was 123 for the lower-uncertainty scenario of the significance framework, 122 for the higher-uncertainty scenario of the significance framework, 128 for the lower-uncertainty scenario of the Bayesian framework, and 144 for the higher-uncertainty scenario of the Bayesian framework. After answering the outcome question, the participants proceeded to two separate pages for the manipulation checks. The participants were not allowed to go back to previous pages in the questionnaire.

\subsection{Treatments}
In the vignette, the participants were shown a hypothetical policymaking scenario, where a group of researchers reports a statistical estimate on the effect of a new bus line on traffic jams. The policy topic is quite neutral and should be fairly free from the ideological dispositions of participants. The neutrality of the topic is important because ``[d]ecision making based on scientific evidence is also affected by differences in perceptions of scientific uncertainty of research fields (and their findings) \textit{as a function of ideology and political affiliation}'' \autocite[287; emphasis added]{Broomell2017}.

There were four treatments: lower uncertainty under the significance framework, higher uncertainty under the significance framework, lower uncertainty under the Bayesian framework, and higher uncertainty under the Bayesian framework. The treatment vignette was as follows:

\begin{quote}
Suppose that you were a policymaker in a government. Suppose also that a group of researchers reported a statistical estimate to you. The report says that if the government introduced a new bus line, it would cost an extra 1\% of the current spending on public transport, but could reduce traffic jams by between 1\%--15\%.

Because this is an estimate, there is some uncertainty around it. In other words, the new bus line might indeed reduce traffic jams as expected, or might have no effect or exacerbate traffic jams. One way to measure the uncertainty is to compute [\textit{something called a p-value. The p-value takes a value between 0\% and 100\%, and if a p-value is smaller than 5\%, it is conventional to conclude that the estimate is ``statistically significant'' and there is evidence for the expected effect}] / [\textit{the probability of an estimate being correct. The probability takes a value between 0\% and 100\%, and a greater value means greater plausibility}].

This time, the estimate is [\textit{\textbf{statistically significant (the p-value is 2\%)}}] / [\textit{\textbf{NOT statistically significant (the p-value is 25\%)}}] / [\textit{\textbf{correct with a probability of 95\%}}] / [\textit{\textbf{correct with a probability of 68\%}}].

Would you, as a policymaker, introduce the new bus line?

Yes\\
No
\end{quote}

\noindent
The italicized texts are information varied among the four treatment groups (while the texts were bold in the vignette as well). The explanation about the \textit{p}-value was shown to the two groups under the significance framework. The one about probability was shown to the two groups under the Bayesian framework. The cost and effect of the new bus line were fixed. 

Using the interval estimate of the effect means that the \textit{p}-values and the Bayesian probabilities were two-tailed. Some participants (especially those who had statistical training previously) might have recalculated them to the one-tailed versions focusing only on the lower bounds and, thereby, understood lower degrees of uncertainty -- the \textit{p}-values to $0.01 = 0.02 / 2$ and $0.125 = 0.25 / 2$, and the probabilities to $0.975 = 0.95 + (1 - 0.95) / 2$ and $0.84 = 0.68 + (1 - 0.68) / 2$, for the effect being equal to, or greater than, 1\%. Even so, the experimental design would remain valid, (1) because the randomization of the treatment assignment was expected to guarantee an equal proportion of participants doing the recalculation per treatment group, and (2) because the quantity of interest was not the difference in the likelihood of adopting the policy within the significance or Bayesian framework but the difference in these within-framework differences (for greater detail, see the section ``Two-Tailed vs. One-Tailed Interpretation'' in the supplemental document).

It might be questioned why I used a \textit{p}-value rather than a frequentist confidence level for the significance framework given the vignette showed the interval estimate. There were two reasons. First, it is rare for researchers to present a frequentist confidence interval other than the one considered as a decision threshold for statistical significance (typically 95\%). Meanwhile, some researchers do display actual \textit{p}-values, even if they use the dichotomy of statistical significance vs. insignificance when making inference. Second, a frequentist confidence interval is often (mis)interpreted as the interval that includes effect sizes with a certain probability (i.e., the Bayesian confidence interval). The Bayesian treatments already cover this (mis)interpretation of frequentist confidence intervals.

The probability of 95\% was chosen for the Bayesian lower-uncertainty treatment, as it is the conventional threshold for an interval estimate to identify statistical significance. The probability of 68\%  for the Bayesian higher-uncertainty treatment was an arbitrary number that was well under the conventional threshold of statistical significance. The \textit{p}-values mathematically corresponding to these Bayesian probabilities were computed for the treatments under the significance framework, based on the assumption that the sampling distribution and the posterior distribution are normal with the same mean and standard deviation. This mathematical correspondence was done to attribute a difference in the outcome (if any) to participants' interpretation of the two uncertainty frameworks (significance vs. Bayesian) only.

The lower-uncertainty treatment under the significance framework used a p-value of 2\%, an approximate two-tailed \textit{p}-value based on the normal distribution with the mean of 8 and the standard deviation of 3.5, which generates the 95\% interval including a value of 1 to 15. The higher-uncertainty treatment employed a p-value of 25\%, an approximate two-tailed \textit{p}-value based on the normal distribution with the mean of 8 and the standard deviation of 7, which generates the 68\% interval including a value of 1 to 15. 

The goal of the analysis was to examine whether participants adopt the policy or not, given the different information of statistical uncertainty, for the same hypothetical policymaking scenario including the same cost and effect size. Although a reduction in traffic jams may well be a socially desirable option (e.g., more economic efficiency and fewer greenhouse gases), whether introducing a new bus line is desirable as a policy depends on how participants evaluate its cost-effectiveness and the uncertainty of the estimate.

Even if there is a variation in a utility function among citizens \autocite[581--82]{Fehr-Duda2012}, provided that the effect size and the cost are fixed and the experiment is randomized, participants should on average be less likely to adopt the policy when they are given the information of greater uncertainty. But also, as both the \textit{p}-value and the Bayesian probability are the continuous measures of statistical uncertainty, the relationship between a decision to adopt the policy and the level of statistical uncertainty should not be deterministic. In other words, there should be more participants not to adopt the policy when provided with a greater degree of statistical uncertainty, but also there should be some participants who have the utility functions that will lead them to adopt the policy even if, for example, the Bayesian probability is 68\%.

The question is whether both the \textit{p}-values coupled with statistical (in)significance and the Bayesian probabilities allow participants to see the continuous nature of uncertainty. Recall that the \textit{p}-values and Bayesian probabilities used here mathematically correspond to each other, under each of the lower vs. higher-uncertainty scenarios. Therefore, it is participants' interpretation of the theoretical framing of uncertainty that should explain the difference (if any) between the lower-uncertainty and higher-uncertainty groups in the likelihood of deciding to introduce the new bus line. The hypothesis is that the significance framework makes people less able to see the continuous nature of uncertainty. Therefore, I expect to find that the difference between the lower-uncertainty and higher-uncertainty groups is greater under the significance framework than under the Bayesian framework. The empirical analysis examines this.

\section{Statistical Modeling}
To analyze the data, I use Bayesian logistic regression. The Markov Chain Monte Carlo software used is Stan \autocite{StanDevelopmentTeam2019b}, implemented via the rstanarm package version 2.21.1 \autocite{Goodrich2020a} on RStudio \autocite{RStudioTeam2020} running R version 4.1.2 \autocite{RCoreTeam2021}.  I use four chains, each of which has 20,000 iterations and then discards the first 1,000 posterior samples. The model is:

\begin{align}\label{model}
\begin{split}
Bus_i &\sim Bernoulli(\text{logit}^{-1}(\pi_{i})),\\
\pi_{i} &=\begin{cases}\beta_0 + \beta_1 LowUncert_i + \beta_2 Bayes_i + \beta_3 Bayes_i \times LowUncert_i \\ \text{ if the demographic controls are not used},\\
\beta_0 + \beta_1 LowUncert_i + \beta_2 Bayes_i + \beta_3 Bayes_i \times LowUncert_i + \mathbf{x}_{i}\boldsymbol{\gamma} \\ \text{ if the demographic controls are used},
\end{cases}
\end{split}
\end{align}

\noindent
where $Bus_i$ is a binary variable coded 0 if a participant $i$ chose ``No'' in the outcome question and 1 if s/he chose ``Yes''; $Bernoulli(\cdot)$ is a Bernoulli distribution; $\text{logit}^{-1}(\cdot)$ is the inverse logit function to map log odds onto probability; $\pi_{i}$ is the log odds of participants deciding to introduce the new bus line; $\beta_0$ is the baseline log odds capturing the treatment category of the higher uncertainty under the significance framework; $\beta_1$ is the log odds ratio coefficient for the variable $LowUncert_i$ coded 1 if a participant was given the information of lower uncertainty and 0 otherwise; $\beta_2$ is the log odds ratio coefficient for the variable $Bayes_i$ coded 1 if a participant was given Bayesian information and 0 otherwise; $\beta_3$ is the log odds ratio coefficient for the product term $Bayes_i \times LowUncert_i$ coded 1 if a participant was given the information of lower uncertainty under the Bayesian framework; and $\mathbf{x}_{i}\boldsymbol{\gamma}$ is the dot product of the vector of the demographic variables $\mathbf{x}_{i}$ (female, other/non-binary, the 25--44, 45--64, and 65+ age groups, a university degree, the previous study or use of statistics) and the vector of the corresponding log odds ratio coefficients $\boldsymbol{\gamma}$. The summary statistics is presented in Table \ref{sumstats}.

\begin{table}[t]
\centering
\begin{tabular}{l c c c c c}
\hline
\hline
Variable & Mean & SD & Min & Max & N \\ 
\hline
Bus	& 0.74 & 0.44 & 0 & 1 & 517\\
Low Uncertainty & 0.49 & 0.50 & 0 & 1 & 517\\
Bayes & 0.53 & 0.50	&0 & 1 & 517\\
Bayes Low Uncertainty & 0.25 & 0.43 & 0 & 1 & 517\\
Female & 0.52 & 0.50 & 0 & 1 & 505\\
Other/Non-Binary & 0.004 & 0.06 & 0 & 1 & 505\\
Aged 18--24 & 0.15 & 0.36 & 0 & 1 & 508\\
Aged 25--44 & 0.37 & 0.48 & 0 & 1 & 508\\
Aged 45--64 & 0.32 & 0.47 & 0 & 1 & 508\\
Aged 65+ & 0.16 & 0.37 & 0 & 1 & 508\\
University Degree & 0.45 & 0.50 & 0 & 1 & 515\\
Statistics & 0.48 & 0.50 & 0 & 1 & 517\\
\hline
\hline
\end{tabular}
\caption{Summary statistics. SD: standard deviation; N: number of observations.}
\label{sumstats}
\end{table}

As for priors, I first use the following weakly informative priors:

\begin{align}
\begin{split}
\beta_0 &\sim N(\mu_{\beta_0}=0, \sigma_{\beta_0}=10),\\
\beta_1 &\sim N(\mu_{\beta_1}=0, \sigma_{\beta_1}=2.5),\\
\beta_2 &\sim N(\mu_{\beta_2}=0, \sigma_{\beta_2}=2.5),\\
\beta_3 &\sim N(\mu_{\beta_3}=0, \sigma_{\beta_3}=2.5),
\end{split}
\end{align}

\noindent
where $N(\cdot)$ is a normal distribution, $\mu$ is the mean, and $\sigma$ is the standard deviation. These weakly informative priors enable data-driven estimation coupled with regularization. The mean of zero for the constant $\beta_0$ makes sense here, as the rstanarm package automatically centers predictors during the estimation process; the returned results are based on the original scales of predictors.\footnote{When the demographic controls are included, $N(\mu=0, \sigma=2.5)$ is also used for each element of $\boldsymbol{\gamma}$.}

As an alternative, I also use the informative priors I theoretically deduced at the preregistration stage. To do so, I disable the automatic centering of predictors in the rstanarm package, so that I can specify the baseline log odds based on the logit function of the probability value I theoretically deduce as follows. The CSO has census data from 2016 on what means of commuting people use \autocite{CentralStatisticsOffice2017}. I computed the proportion of the commuters (19+ years old students or 15+ years old workers) who use buses, minibuses, coaches, cars (as a driver or as a passenger), vans, and lorries, to the total number of the respondents in the census, which is approximately 76\%.\footnote{In the census data, lorries are included in the category ``Other,'' which accounts for 0.4\% of the total responses. Whether I included this category into the calculation or not, the total proportion remains approximately 76\%.} I assume these people are concerned with traffic jams because they are directly affected by traffic jams and, therefore, should most appreciate the positive implications of fewer traffic jams for the society. The remaining categories of the means of commuting are walking, bicycles, trains, trams, motorcycles, scooters, work at/from home, and no answer.\footnote{The proportion of no answers in the census data is 3\%.}

Based on these points, I set the informative priors as follows:

\begin{align}\label{infPrior}
\begin{split}
\beta_0 &\sim N(\mu_{\beta_0}=\text{logit}(0.07), \sigma_{\beta_0}=1.5),\\
\beta_1 &\sim N(\mu_{\beta_1}=\text{logit}(0.76) - \mu_{\beta_0}, \sigma_{\beta_1}=0.25),\\
\beta_2 &\sim N(\mu_{\beta_2}=\text{logit}(0.76 \times 0.50) - \mu_{\beta_0}, \sigma_{\beta_2}=0.25),\\
\beta_3 &\sim N(\mu_{\beta_3}=\text{logit}(0.76 \times 0.95) - \mu_{\beta_0} - \mu_{\beta_1} - \mu_{\beta_2}, \sigma_{\beta_3}=0.25).
\end{split}
\end{align}

\noindent
The prior predictive distributions (the distribution of predicted outcome values given the model without the data) based on these informative priors are available in the supplemental document.

The reason another/other $\beta$ values are subtracted for the mean of all but the prior for $\beta_0$ in equation \ref{infPrior} is algebraic. Equation \ref{model} implies:

\begin{align}
\begin{split}
\pi _{i} &=
\begin{cases}\beta_0 & \text{ if } LowUncert_i=0 \text{ and } Bayes_i=0,\\
\beta_0 + \beta_1 & \text{ if } LowUncert_i=1 \text{ and } Bayes_i=0,\\
\beta_0 + \beta_2 & \text{ if }  LowUncert_i=0 \text{ and } Bayes_i=1,\\
\beta_0 + \beta_1 + \beta_2 + \beta_3 & \text{ if }  LowUncert_i=1 \text{ and } Bayes_i=1.
\end{cases}
\end{split}
\end{align}

\noindent
Since $\pi$ is the log odds of participants deciding to introduce the new bus line, then a value for a particular $\beta_j$ is $\pi$ minus all the remaining terms. For example, for the Bayesian low-uncertainty case (i.e., $LowUncert_i=1$ \& $Bayes_i=1$), $\beta_3 = \pi_{i} - \beta_0 - \beta_1 - \beta_2$.

The rationale for choosing the above values for the logit functions is as follows. To avoid confusion, hereafter I use ``likelihood'' to denote the probability of the outcome variable taking a value of one, while ``probability'' to refer to the probability that a causal factor has an effect on the outcome.

$\text{logit}(0.07)$ is used for the mean of the prior for $\beta_0$, as I assume even those who are concerned with traffic jams would interpret statistical insignificance as evidence for the absence of the effect. The number, .07, is the proportion of bus, minibus, and coach users in the census data; I assume they might be most interested in the effect of the new bus line from their direct experience of using buses, regardless of its effect being certain or uncertain. However, as I am not confident of this reasoning, I put a large standard deviation of 1.5 for this prior. This makes the prior predicted likelihood of participants deciding to introduce the new bus line vary between 0.004 and 0.6 within +/- two standard deviations. Thus, the prior is only weakly informative.

I have stronger priors for the remaining $\beta$s. For $\beta_1$, the likelihood of 0.76 is chosen, as I assume that almost all of those who are concerned with traffic jams will decide to introduce the new bus line, given the statistical significance of the effect. A standard deviation of 0.25 then allows the prior predicted likelihood of participants deciding to do so to vary between 0.66 and 0.84 within +/- two standard deviations.

For $\beta_2$, the likelihood of 0.76 is multiplied by 0.50, which results in a likelihood of 0.38. In other words, I assume that a half of those who are concerned with traffic jams will be risk-tolerant enough or have preference great enough to decide to introduce the new bus line, even if the probability they are informed of in the vignette is the relatively low 68\%. A standard deviation of 0.25 then allows the prior predicted likelihood of participants deciding to do so to vary between 0.27 and 0.50 within +/- two standard deviations.

Finally, for $\beta_3$, the probability of 0.76 is multiplied by 0.95, which results in a likelihood of 0.72. In other words, I expect that most of those who are concerned with traffic jams will decide to introduce the new bus line when the probability they are informed of in the vignette is the near certainty of 95\%. A standard deviation of 0.25 then allows the prior predicted likelihood of participants deciding to do so to vary between 0.61 and 0.81 within +/- two standard deviations.

The quantity of interest is:

\begin{equation}
DD = \{\underbrace{(\beta_0 + \beta_1) - \beta_0}_{\text{difference under the significance framework}}\} - \{\underbrace{(\beta_0 + \beta_1 + \beta_2 + \beta_3) - (\beta_0 + \beta_2)}_{\text{difference under the Bayesian framework}}\},
\end{equation}

\noindent
where $DD$ stands for the difference in the differences between the groups under the significance framework and those under the Bayesian framework.\footnote{To be clear, the difference in the differences here has nothing to do with the so-called difference-in-differences estimator for panel data.} If the significance framework makes participants less able to see the continuous nature of uncertainty than the Bayesian framework, $DD$ should be a positive value. I estimate the posterior predicted values of $DD$ based on posterior samples from the Bayesian logistic regression.\footnote{Where the demographic covariates are included in the regression model, I take the average over all $\text{logit}^{-1}(\pi_{i})$.}

\section{Results}
The $\hat{R}$ was approximately 1.00 for every estimated parameter, suggesting the models did not fail to converge. The effective sample size exceeded at least 32,000 for every estimated parameter. The posterior predictive checks imply the models fit data well (see the supplemental document). The posteriors of the log odds ratio coefficients are available in the supplemental document.

\begin{figure}[t]
	\includegraphics[scale=0.175]{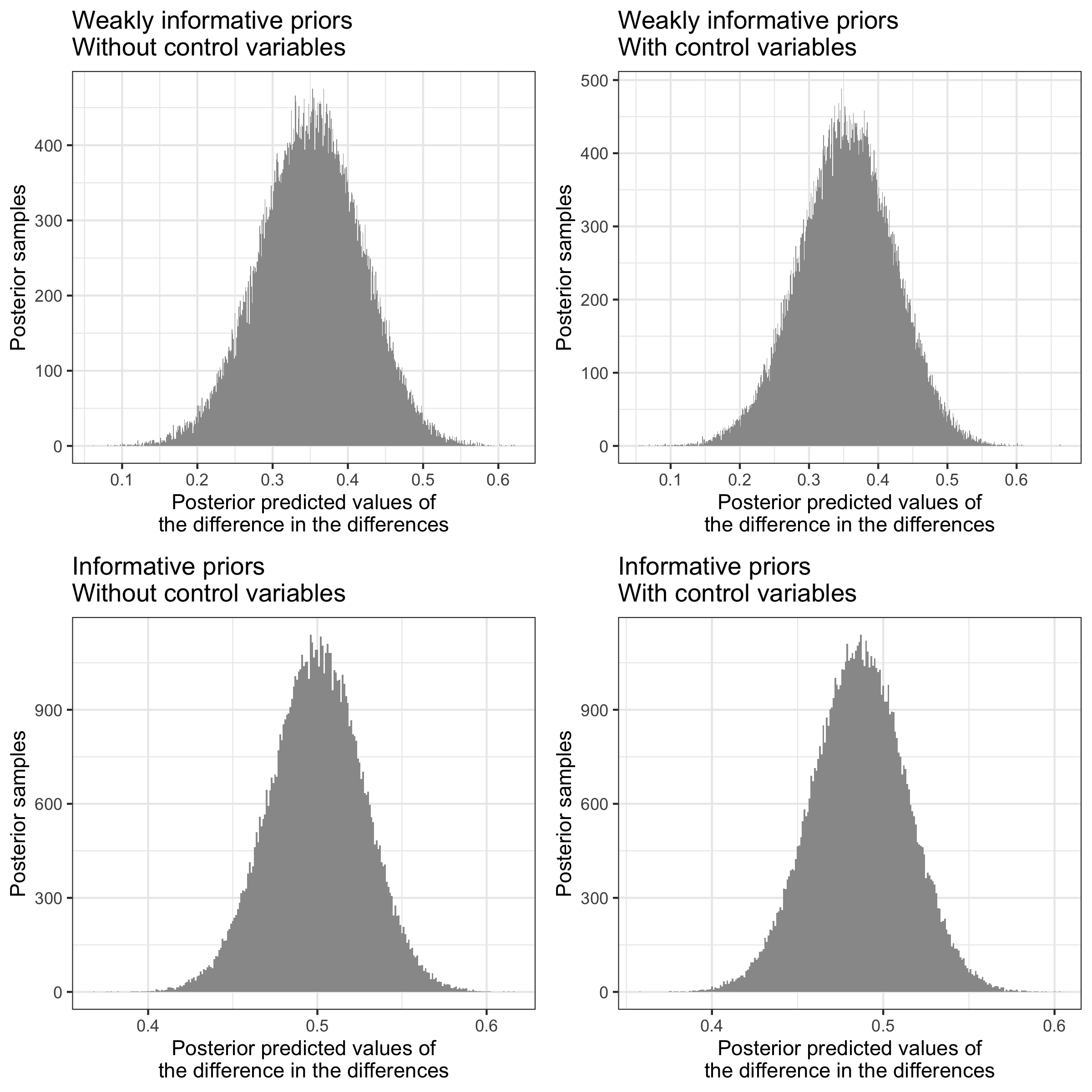}
	\centering
	\caption{Posterior predicted values of the difference in the differences.}
	\label{postPredPlots}
\end{figure}

The posterior predicted values of the difference in the differences is plotted in Figure \ref{postPredPlots}, for each of the models with/without the control variables and with the weakly informative priors or the informative priors.\footnote{The data visualization was done by the ggplot2 package \autocite{Wickham2016}.} It is clear from the plots that in every model all predicted values are positive. The variance is much smaller when the informative priors are used than when the weakly informative priors are used; there is no clear difference between the models with and without the control variables given the same set of priors.

\begin{figure}[t]
	\includegraphics[scale=0.175]{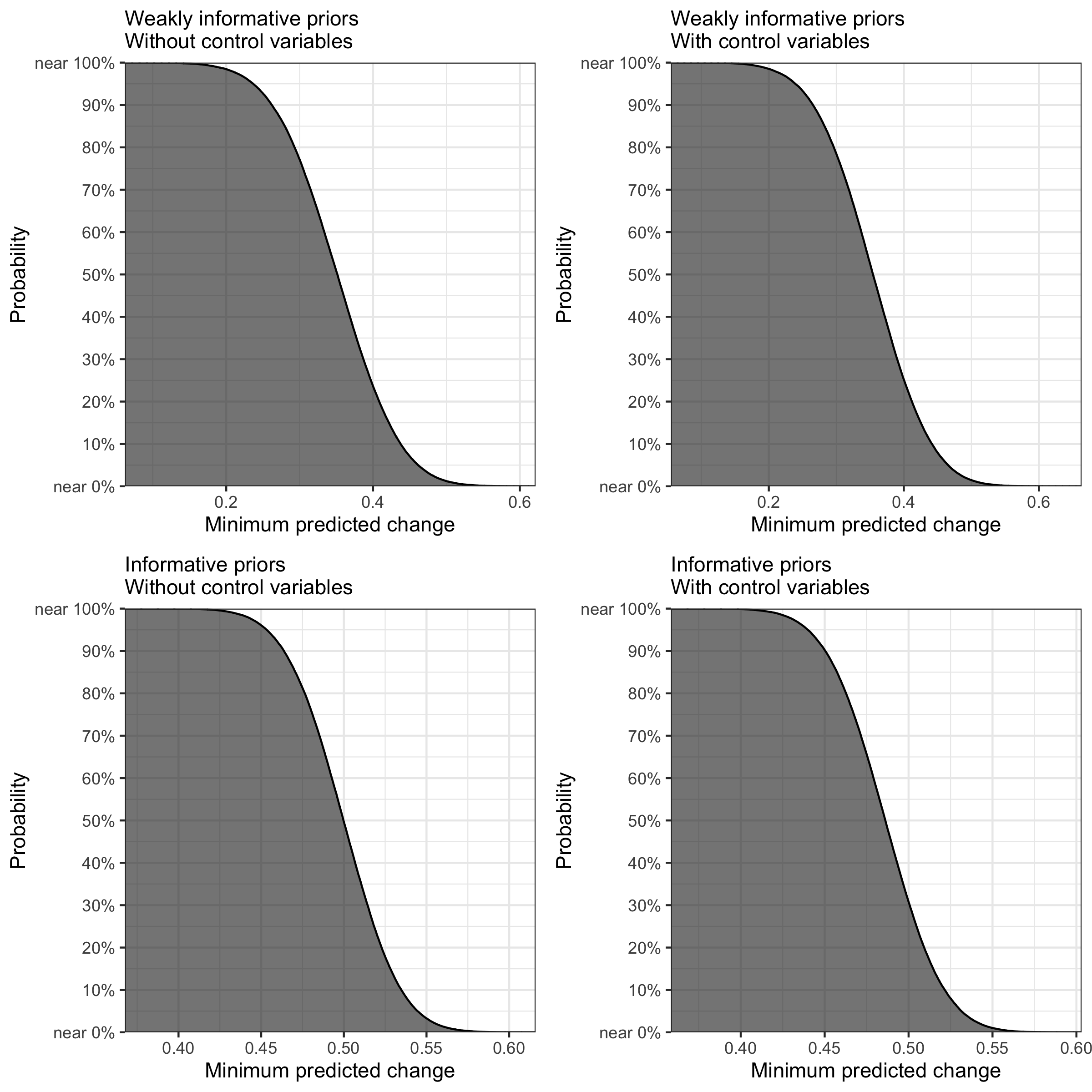}
	\centering
	\caption{Probability of the difference in the differences.}
	\label{ccdfPlots}
\end{figure}

Figure \ref{ccdfPlots} presents the probability that the difference in the differences has at least a certain size, based on a complementary cumulative distribution function \autocite{Suzuki2022}. When the weakly informative priors are used, the difference in the differences is near 100\% probable to be at least 0.15. In other words, it is near 100\% probable, given the data and model assumptions, that the difference in the likelihood of participants accepting the bus line is 0.15 points greater when they are under the significance framework than under the Bayesian framework. When the informative priors are used, the effect size is larger. It is near 100\% probable, given the data and model assumptions, that the difference in the likelihood of participants accepting the bus line is 0.40 points greater when they are under the significance framework than under the Bayesian framework

These results suggest, as expected, that the significance framework of uncertainty results in a greater difference between the lower-uncertainty and higher-uncertainty treatments in the likelihood of accepting the proposed bus line, than the Bayesian framework. Recall that the p-value and the Bayesian probability per degree of uncertainty (lower vs. higher) were defined so that they mathematically corresponded to each other. This suggest that the difference must be driven by the different theoretical framing of these mathematically corresponding quantities. Under their lower-uncertainty scenarios, both frameworks produce the posterior predictions of similarly high predicted likelihood of participants choosing to adopt the policy (see section 7 in the supplemental documents for the plots).\footnote{One puzzling finding is that the proportion of those deciding to adopt the policy was larger in the group shown lower uncertainty under the Bayesian framework than in the group shown lower uncertainty under the significance framework (0.91 against 0.82). It is unclear exactly why, apart from a random chance. Future research could try to replicate this finding. This point, however, does not affect the main conclusion that the Bayesian framework enabled participants to see the continuous nature of uncertainty more clearly than the significance framework.} This implies the significance framework was prone to leading the participants to (mis)understand statistical insignificance is evidence for the absence of an effect.

It might be argued that actually the significance framework enabled participants to see the continuous nature of uncertainty more clearly than the Bayesian framework. However, such an argument is inconsistent with the existing behavioral literature that has found that a continuous, numerical probability scale enables people to see and use the measure of uncertainty more accurately than a discrete, verbal scale  \autocites[e.g.,][]{Mandel2021, McShane2016, McShane2017, Mislavsky2021, Friedman2018}.

In addition, my data are more consistent with the idea that the Bayesian framework enabled participants to see the continuous nature of uncertainty more clearly. Recall my reasoning for the informative priors. I \textit{a priori} expected the mean likelihood of $0.76$ under the lower-uncertainty scenario of the significance framework, $0.07$ under the higher-uncertainty scenario of the significance framework, $0.72 = 0.76 \times 0.95$ under the lower-uncertainty scenario of the Bayesian framework, and $0.38 = 0.76 \times 0.50$ under the higher-uncertainty scenario of the Bayesian framework. The actual observed mean likelihood (i.e., the proportion of the ``Yes'' answers to the sum of ``Yes'' and ``No'' answers to the outcome question per treatment group) is $0.82 = 101 / (101 + 22)$, $0.39 = 48 / (48  + 74)$, $0.91 = 116 / (116 + 12)$, and $0.83 = 120 / (120 + 24)$, respectively. This means that my prior reasoning for the proportion of ``Yes'' was an underestimate in \textit{every group}. It might be because those commuters who are not affected by traffic jams were still interested in reducing them, for example, for economic or environmental reasons, and/or because a greater proportion of those who are affected by traffic jams were risk-acceptant people than the initially assumed 0.50.

Regardless of the reasons why my \textit{a priori} expectations were underestimates, what is informative here is that the degree of my underestimation is much greater in the groups under the higher-uncertainty scenarios than in the groups under the lower-uncertainty scenarios. The observed mean likelihood minus my \textit{a priori} estimated mean likelihood is $0.06$ under the lower-uncertainty scenario of the significance framework, $0.32$ under the higher-uncertainty scenario of the significance framework, $0.18$ under the lower-uncertainty scenario of the Bayesian framework, and $0.45$ under the higher-uncertainty scenario of the Bayesian framework. This suggests there were more participants than I expected \textit{a priori} who valued the importance of the new bus line and a reduction in traffic jams, so much that they were risk-acceptant enough to adopt the policy, even under the higher-uncertainty conditions. If so, the expected proportion of ``Yes'' in the higher-uncertainty group should be significantly greater, than my originally expected proportion of 0.38 (which is the percentage of people who should be concerned with traffic jams, 0.76, times my original expectation of the percentage of the people who are risk-acceptant enough to adopt the policy under the higher-uncertainty scenario, 0.50).

This is the case only under the higher-uncertainty scenario of the Bayesian framework ($0.83$) and not under the higher-uncertainty scenario of the significance framework ($0.39$). This implies that although participants were risk-acceptant enough to adopt the policy even under the higher level of uncertainty, the information of higher-uncertainty under the significance framework discouraged them from doing so. Put differently, the finding suggests that participants told about statistical insignificance were more likely to see it as evidence for the null effect. For thee reasons, I infer that the significance framework was less likely to enable the participants to see the continuous nature of uncertainty, than the Bayesian framework.

In the supplemental document, I also present the results from models using gender, age groups, a university degree, or the previous experience of statistics as an effect modifier. This is an exploratory analysis because I did not theorize or plan it at the pre-registration stage but realized it might be informative for future research. The detailed description of the results are available in the supplemental document. To summarize them, there is modest evidence that those who are female, 45+ years old, and have a university degree, on average, exhibited a greater difference in adopting or not adopting the policy under the significance framework than under the Bayesian framework. One possible explanation is that these participants were more risk-averse.

Meanwhile, I found there was very weak evidence that those who previously did NOT study or use statistics, on average, exhibited a greater difference in adopting or not adopting the policy under the significance framework than under the Bayesian framework. This finding contrasts with the finding in \textcite{McShane2016} that, when experiment participants were asked to describe (rather than make inference over) a difference in life expectancy between two different treatment groups, statistically untrained undergraduate students were on average unaffected by the threshold of a \textit{p}-value of 5\%. This difference might be attributable to whether or not participants received an explanation about a \textit{p}-value and the conventional threshold of $p<5\%$ for statistical significance. In \textcite{McShane2016}, the participants did not, while in my experiment the participants did receive a brief explanation of these. It might be the case that once statistically untrained people are informed about the convention of statistical significance, they are at least as prone to see statistical significance as evidence for an effect and statistical insignificance as evidence for the lack of an effect, as those who are statistically trained. And if some of statistically trained people knew the controversy over \textit{p}-values, statistically trained people might, on average, be less prone to such a tendency than statistically untrained people who would have been informed only about the convention of statistical significance.

\section{Conclusion}
Using an original survey experiment, this article has examined which type of statistical uncertainty -- statistical (in)significance with a \textit{p}-value, or a Bayesian probability -- helps people see the continuous nature of uncertainty and make a policy decision accordingly. The participants read a hypothetical policymaking situation to decide whether to introduce a new bus line to reduce traffic jams, given its cost, effect size, and uncertainty level.

I used the following four treatments:  statistical significance with a \textit{p}-value of 2\%, statistical insignificance with a \textit{p}-value of 25\%, a 95\% Bayesian probability, and a 68\% Bayesian probability. These uncertainty quantities were defined so that they mathematically corresponded to each other per degree of uncertainty (lower vs. higher). If the participants had been able to see the continuous nature of uncertainty, we should have observed little difference between the significance framework and the Bayesian framework, with respect to the difference between the lower-uncertainty and higher-uncertainty treatments in the likelihood of participants deciding to introduce the new bus line. It turned out that this was not the case. The significance framework resulted in a greater difference between the lower-uncertainty and higher-uncertainty scenarios, than the Bayesian framework. Overall, the findings suggest the Bayesian framework was more likely to enable the participants to see the continuous nature of uncertainty than the significance framework.

The results of this article suggest that presenting a Bayesian probability is a more effective way to communicate the continuous measure of uncertainty over whether a policy achieves a desirable outcome, for evidence-based policymaking. Just presenting an actual \textit{p}-value alongside statistical (in)significance does not seem to remedy people's tendency to deterministically think statistical significance means evidence for an effect while statistical insignificance means evidence for the lack of an effect, if those who receive the information know, or are told about, the convention of statistical significance.

Future research might examine whether presenting only a \textit{p}-value will produce the same outcomes as when a Bayesian probability is used. Another point for future research is whether a Bayesian probability will also fail to enable people to see the continuous nature of uncertainty, if it is used together with statistical significance (e.g., a 95\% probability used as a threshold for statistical significance). Although the dichotomy of statistical significance vs. insignificance is dominant in applied research using statistics, this article suggests that it may hinder the appreciation of the continuous nature of uncertainty not only among researchers but also among the public.

\section*{Acknowledgments}
I would like to thank Johan A. Dornschneider-Elkink for his helpful comments. I would like to acknowledge the receipt of funding from the Irish Research Council (the grant number: GOIPD/2018/328) for the development of this work. The views expressed are my own unless otherwise stated, and do not necessarily represent those of the institutes/organizations to which I am/have been related.

\section*{Supplemental Materials}
The supplemental document and the replication package are available on my website at \url{https://akisatosuzuki.github.io}.

\printbibliography

@article{Broomell2017,
  title = {Public Perception and Communication of Scientific Uncertainty},
  author = {Broomell, Stephen B. and Kane, Patrick Bodilly},
  date = {2017},
  journaltitle = {Journal of Experimental Psychology: General},
  volume = {146},
  number = {2},
  pages = {286--304}
}

@online{CenterforOpenScience2020,
  title = {{{OSF Registries}}},
  author = {{Center for Open Science}},
  date = {2020},
  url = {https://osf.io/registries},
  urldate = {2020-10-21}
}

@online{CentralStatisticsOffice2017,
  title = {Population {{Usually Resident}} and {{Present}} in the {{State}} 2011 to 2016 by {{Sex}}, {{Nationality}}, {{CensusYear}}, {{At Work School}} or {{College}} and {{Means}} of {{Travel}}},
  author = {{Central Statistics Office}},
  date = {2017},
  url = {https://data.cso.ie/},
  urldate = {2020-09-08}
}

@online{CentralStatisticsOffice2020,
  title = {Information {{Society Statistics}} - {{Households}} 2019: {{Household Internet Connectivity}}},
  author = {{Central Statistics Office}},
  date = {2020},
  url = {https://www.cso.ie/en/releasesandpublications/ep/p-isshh/informationsocietystatistics-households2019/householdinternetconnectivity/},
  urldate = {2020-09-08}
}

@online{CentralStatisticsOffice2021,
  title = {Population {{Estimates}} by {{Age Group}} and {{Sex}}},
  author = {{Central Statistics Office}},
  date = {2021},
  url = {https://data.cso.ie/},
  urldate = {2021-09-20}
}

@article{Fehr-Duda2012,
  title = {Probability and {{Risk}}: {{Foundations}} and {{Economic Implications}} of {{Probability-Dependent Risk Preferences}}},
  author = {Fehr-Duda, Helga and Epper, Thomas},
  date = {2012},
  journaltitle = {Annual Review of Economics},
  volume = {4},
  pages = {567--593}
}

@article{Friedman2018,
  title = {The {{Value}} of {{Precision}} in {{Probability Assessment}}: {{Evidence}} from a {{Large-Scale Geopolitical Forecasting Tournament}}},
  author = {Friedman, Jeffrey A. and Baker, Joshua D. and Mellers, Barbara A. and Tetlock, Philip E. and Zeckhauser, Richard},
  date = {2018},
  journaltitle = {International Studies Quarterly},
  volume = {62},
  number = {2},
  pages = {410--422}
}

@unpublished{Goodrich2020a,
  title = {{{rstanarm}}: {{Bayesian}} Applied Regression Modeling via {{Stan}}},
  author = {Goodrich, Ben and Gabry, Jonah and Ali, Imad and Brilleman, Sam},
  date = {2020},
  url = {https://mc-stan.org/rstanarm},
  howpublished = {R package version 2.21.1}
}

@article{Kruschke2018,
  title = {The {{Bayesian New Statistics}}: {{Hypothesis}} Testing, Estimation, Meta-Analysis, and Power Analysis from a {{Bayesian}} Perspective},
  author = {Kruschke, John K. and Liddell, Torrin M.},
  date = {2018},
  journaltitle = {Psychonomic Bulletin and Review},
  volume = {25},
  number = {1},
  pages = {178--206}
}

@article{Lew2012,
  title = {Bad Statistical Practice in Pharmacology (and Other Basic Biomedical Disciplines): {{You}} Probably Don't Know {{P}}},
  author = {Lew, Michael J.},
  date = {2012},
  journaltitle = {British Journal of Pharmacology},
  volume = {166},
  number = {5},
  pages = {1559--1567}
}

@article{Mandel2021,
  title = {Facilitating {{Sender-Receiver Agreement}} in {{Communicated Probabilities}}: {{Is It Best}} to {{Use Words}}, {{Numbers}} or {{Both}}?},
  author = {Mandel, David R. and Irwin, Daniel},
  date = {2021},
  journaltitle = {Judgment and Decision Making},
  volume = {16},
  number = {2},
  pages = {363--393}
}

@article{McShane2016,
  title = {Blinding {{Us}} to the {{Obvious}}? {{The Effect}} of {{Statistical Training}} on the {{Evaluation}} of {{Evidence}}},
  author = {McShane, Blakeley B. and Gal, David},
  date = {2016},
  journaltitle = {Management Science},
  volume = {62},
  number = {6},
  pages = {1707--1718}
}

@article{McShane2017,
  title = {Statistical {{Significance}} and the {{Dichotomization}} of {{Evidence}}},
  author = {McShane, Blakeley B. and Gal, David},
  date = {2017},
  journaltitle = {Journal of the American Statistical Association},
  volume = {112},
  number = {519},
  pages = {885--895}
}

@article{Mislavsky2021,
  title = {Combining {{Probability Forecasts}}: 60\% and 60\% {{Is}} 60\%, but {{Likely}} and {{Likely Is Very Likely}}},
  shorttitle = {Combining {{Probability Forecasts}}},
  author = {Mislavsky, Robert and Gaertig, Celia},
  date = {2021},
  journaltitle = {Management Science},
  shortjournal = {Management Science},
  pages = {1--23},
  issn = {0025-1909, 1526-5501},
  doi = {10.1287/mnsc.2020.3902},
  langid = {english}
}

@article{Montgomery2018b,
  title = {How {{Conditioning}} on {{Posttreatment Variables Can Ruin Your Experiment}} and {{What}} to {{Do}} about {{It}}},
  author = {Montgomery, Jacob M. and Nyhan, Brendan and Torres, Michelle},
  date = {2018},
  journaltitle = {American Journal of Political Science},
  shortjournal = {American Journal of Political Science},
  volume = {62},
  number = {3},
  pages = {760--775},
  langid = {english}
}

@online{OECD2019,
  title = {Governing Better through Evidence-Informed Policy Making},
  author = {{OECD}},
  date = {2019},
  url = {http://www.oecd.org/gov/governing-better-through-evidence-informed-policy-making.htm},
  urldate = {2019-08-16}
}

@report{RCoreTeam2021,
  title = {R: {{A Language}} and {{Environment}} for {{Statistical Computing}}},
  author = {{R Core Team}},
  date = {2021},
  institution = {{R Foundation for Statistical Computing}},
  location = {{Vienna, Austria}},
  url = {https://www.R-project.org}
}

@report{RStudioTeam2020,
  title = {{{RStudio}}: {{Integrated Development}} for {{R}}},
  author = {{RStudio Team}},
  date = {2020},
  location = {{RStudio, PBC, Boston, MA}},
  url = {http://www.rstudio.com/}
}

@book{ScienceAdviceforPolicybyEuropeanAcademies2019,
  title = {Making Sense of Science for Policy under Conditions of Complexity and Uncertainty},
  author = {{Science Advice for Policy by European Academies}},
  date = {2019},
  publisher = {{SAPEA}},
  location = {{Berlin}},
  url = {https://doi.org/10.26356/MASOS}
}

@misc{StanDevelopmentTeam2019b,
  title = {Stan {{Reference Manual}}, {{Version}} 2.27},
  author = {{Stan Development Team}},
  date = {2019},
  url = {https://mc-stan.org/docs/2_27/reference-manual/index.html}
}

@online{Suzuki2020e,
  title = {Which {{Type}} of {{Statistical Uncertainty Helps Evidence-Based Policymaking}}? {{An Insight}} from a {{Survey Experiment}} in {{Ireland}}},
  author = {Suzuki, Akisato},
  date = {2020-08-05},
  url = {https://osf.io/mgh4v},
  organization = {{OSF Registries}}
}

@unpublished{Suzuki2022,
  title = {Presenting the {{Probabilities}} of {{Different Effect Sizes}}: {{Towards}} a {{Better Understanding}} and {{Communication}} of {{Statistical Uncertainty}}},
  author = {Suzuki, Akisato},
  date = {2022},
  location = {{arXiv: 2008.07478v4 [stat.AP]}},
  url = {https://arxiv.org/abs/2008.07478}
}

@article{Visschers2009,
  title = {Probability Information in Risk Communication: {{A}} Review of the Research Literature},
  author = {Visschers, Vivianne H.M. and Meertens, Ree M. and Passchier, Wim W.F. and De Vries, Nanne N.K.},
  date = {2009},
  journaltitle = {Risk Analysis},
  volume = {29},
  number = {2},
  pages = {267--287}
}

@book{Wickham2016,
  title = {{{ggplot2}}: {{Elegant Graphics}} for {{Data Analysis}}},
  author = {Wickham, Hadley},
  date = {2016},
  edition = {2},
  publisher = {{Springer International Publishing}},
  location = {{Cham}}
}

\end{document}